\documentclass[10pt,twocolumn,reprint,amsmath,amssymb,aps,prx,showpacs,longbibliography,superscriptaddress]{revtex4-1}
\usepackage[latin9]{inputenc}
\usepackage{amsmath}
\usepackage{amssymb}
\usepackage{graphicx}

\makeatletter

\usepackage{amsfonts}
\usepackage{graphicx}
\usepackage{graphics}

\makeatother

\begin{document}
\title{Exact Polaron-Polaron interactions in a Quantum Hall Fluid }
\author{Jia Wang}
\affiliation{Centre for Quantum Technology Theory, Swinburne University of Technology,
Melbourne 3122, Australia}
\author{Xia-Ji Liu}
\affiliation{Centre for Quantum Technology Theory, Swinburne University of Technology,
Melbourne 3122, Australia}
\author{Hui Hu}
\affiliation{Centre for Quantum Technology Theory, Swinburne University of Technology,
Melbourne 3122, Australia}
\date{\today}
\begin{abstract}
We present an exact solution for effective polaron-polaron interactions
between heavy impurities, mediated by a sea of non-interacting light
fermions in the quantum Hall regime with highly degenerate Landau
levels. For weak attraction between impurities and fermions, where
only the manifold of lowest Landau levels is relevant, we obtain an
analytical expression of mediated polaron-polaorn interactions. Remarkably,
polaron interactions are exactly zero when fermions in lowest Landau
levels outnumber heavy impurities. For strong attraction, different
manifolds of higher Landau levels come into play and we derive a set
of equations that can be used to numerically solve the mediated polaron
interaction potential. We find that the potential vanishes when the
distance $R$ between impurities is larger than the magnetic length,
but strongly diverges at short range following a Coulomb form $-1/R$.
Our exact results of polaron-polaron interactions might be examined
in cold-atom setups, where a system of Fermi polarons in the quantum
Hall regime is realized with synthetic gauge field or under fast rotation.
Our predictions could also be useful to understand the effective interaction
between exciton-polarons in electron-doped semiconductors under strong
magnetic field. 
\end{abstract}
\maketitle
\textbf{Introduction} - The exchange of particles leading to effective
interactions is a profound conceptual advance in physics. Fundamentally,
forces mediated by gauge bosons explain all interactions between elementary
particles \citep{Weinberg1995book,Hooft1980SA,Leike1999PR}. In superconductors,
Cooper pairing is induced by exchanging phonons between electrons
\citep{Tinkham1975book}. The nature of effective interactions depends
on the dispersion relation of exchanged particles. For instance, relativistic
massive bosons induce Yukawa potentials \citep{Yukawa1935,Yukawa1937},
and non-relativistic fermions with parabolic dispersion mediate Ruderman--Kittel--Kasuya--Yosida
(RKKY) interactions \citep{Ruderman1954PR,Kasuya1956PoTP,Yosida1957PR}.
Recent developments in engineering quantum many-body systems make
the investigation of these mediated interactions experimentally accessible
\citep{Jia2015PRL,Chin2019Nature,Muir2022,Bruun2024NatPhys,Bruun2024arXiv},
in the previously untouched territory of strong coupling. Specific
interests are paid to the situations where the number of exchange
particles is large, such as in a filled Fermi sea of fermions, so
the whole system can be well described as a polaron problem \citep{Alexandrov2010book},
and the mediated interactions can be regarded as polaron-polaron interactions.
In controllable cold-atom experiments, great efforts have been taken
to measure fermion-mediated polaron interactions by immersing fewer
impurity atoms into a Fermi sea \citep{Chin2019Nature,Bruun2024NatPhys,Bruun2024arXiv}.
In two-dimensional materials such as WS$_{2}$, interaction effects
between Fermi polarons have also been observed \citep{Muir2022}.
However, accurate theoretical descriptions of polaron-polaron interactions
are notoriously difficult to establish \citep{Bruun2024NatPhys,Bruun2024arXiv},
particularly in the strong correlated regime beyond the perturbative
RKKY paradigm.

In this Letter, we aim to explore an intriguing question: what happens
to polaron interactions if the energy spectrum of exchange particles
is highly degenerate? Energy degeneracy, where distinct quantum states
share the same energy, is ubiquitous in quantum mechanics. The presence
of degenerate energy levels typically indicates symmetry in the corresponding
Hamiltonian, with higher degrees of symmetry leading to greater degeneracy.
Notable examples include highly excited states in a spherical harmonic
trap and Rydberg electronic states in a Coulomb potential \citep{Sakurai2017book}.
A highly excited Rydberg electron can mediate an exotic trilobite
potential, binding another ground state atom into an ultra-long-range
molecule \citep{Greene2000PRL,Greene2001PRA}.

Here, we focus on a prime example of Landau levels (LL) of fermions
such as electrons moving in two dimensions (2D) under a perpendicular
magnetic field, which have a macroscopic degeneracy. These highly
degenerate LLs play a crucial role in understanding the integer and
fractional quantum Hall effects, known for the insensitivity of the
quantization of Hall conductivity to material details, such as impurities
\citep{Laughlin1983PRL,Laughlin1999RMP}. The presence of impurities
in LLs has attracted strong interest due to their crucial role in
understanding the robustness of the quantum Hall effect. For example,
short-range impurities can induce significant level broadening effects
\citep{Wegner1983ZPB,Pule1997JSP,Pule1999CMMP}. However, the property
modification of impurities from the polaron perspective and the associated
induced polaron interactions between them in the presence of scattering
particles with high energy degeneracy, such as LLs, have not been
thoroughly considered \citep{Efimkin2018PRB}.

Remarkably, we find that polaron-polaron interactions with Landau
levels are exactly solvable when impurities are infinitely heavy.
This situation arises in both cold-atom setups and 2D semiconductor
experiments, where the mass ratio of impurities to fermions in the
Fermi sea is selectable, so the heavy impurity limit can be approximately
realized. In cold-atom setups, the quantum Hall regime is within reach
either with strong synthetic gauge field \citep{Spielman2009Nature,Zhou2023Science}
or under fast rotation \citep{CornellPRL2004,Zwierlein2021Science,Zwierlein2022Nature}.
Experimentally, exciton-polariton-polarons in GaAs quantum wells were
already studied via polariton spectroscopy in the integer and fractional
quantum Hall regimes \citep{Imamoglu2018PRL}. Our exact and sometime
analytical solutions for polaron-polaron interactions, although restricted
to the heavy polaron limit, offer crucial insights into fundamental
principles of fermion-mediated interactions in the strongly correlated
regime and serve as benchmarks for various approximations, which lead
to debatable predictions on polaron-polaron interactions so far \citep{Bruun2024NatPhys,Bruun2024arXiv}.

Our system consists of a quantum mixture of heavy impurities and light
fermions, where the light fermions are governed by a Hamiltonian with
highly degenerate single-particle energy levels, such as LLs, and
with negligible intraspecies interactions. In a Born-Oppenheimer approximation
or in the rigorous infinitely-heavy limit achievable by confinement
in a deep optical trap or an optical tweezer, the heavy impurities
are localized. The Hamiltonian for $S$ impurities located at positions
$\mathbf{r}_{s}$ is thus given by $H=T+V\equiv T+\sum_{s=1}^{S}g_{s}\delta(\mathbf{r}-\mathbf{r}_{s})$,
where $T$ is a non-interacting Hamiltonian describing certain highly
degenerate manifolds. Here, $\mathbf{r}$ is the position operator
for the light fermions and $g_{s}$ is the coupling strength. We focus
on attractive coupling $g_{s}<0$ and leave the discussions of the
repulsive coupling cases to Appendix \ref{sec:App1}. In the following,
we always discuss the general situation of highly degenerate levels
first and then consider LLs as a concrete example. 

\textbf{Weak impurity-fermion coupling} - We start by investigating
the simplest case of a single weakly interacting impurity, where the
interaction term reduces to $g_{1}\delta(\mathbf{r}-\mathbf{r}_{1})$,
and $g_{1}$ is weak enough so that only the lowest degenerate manifold
with $D$-fold degeneracy is relevant. Expanding the Hamiltonian with
the degenerate eigenfunctions leads to a matrix form with elements
$H_{nn^{\prime}}=T_{nn^{\prime}}+V_{nn^{\prime}}$, where $T_{nn^{\prime}}=\varepsilon\delta_{nn^{\prime}}$
and $V_{nn^{\prime}}=g_{1}\sum_{n,n^{\prime}=1}^{D}\phi_{n}^{*}(\mathbf{r}_{1})\phi_{n^{\prime}}(\mathbf{r}_{1})$.
Here $\phi_{n}^{*}(\vec{r})$ are the eigenstates with degenerate
energy $\varepsilon$. To solve the Hamiltonian, we only need to diagonalize
$V$, as $T$ is proportional to an identity matrix. A crucial observation
is that $V$ is a rank-$1$ matrix and can be expressed as $V=g_{1}\mathcal{N}_{1}\vec{v}_{1}\vec{v}_{1}^{\dagger}$,
where $\vec{v}_{1}^{\dagger}=\left[\phi_{1}(\mathbf{r}_{1}),\phi_{2}(\mathbf{r}_{1}),\cdots,\phi_{D}(\mathbf{r}_{1})\right]/\sqrt{\mathcal{N}_{1}}$
is a row vector, and $\mathcal{N}_{1}=\sum_{n=1}^{D}\left|\phi_{n}(\mathbf{r}_{1})\right|^{2}$
can be regarded as a normalization constant. The diagonalization can
then be obtained analytically as $V\vec{v}_{1}=g_{1}\mathcal{N}_{1}\vec{v}_{1},$
making $\vec{v}_{1}$ an eigenvector with eigenvalue $g_{1}\mathcal{N}_{1}$.
The corresponding real-space wave-function is given by $\psi_{1}(\mathbf{r})=\sum_{n=1}^{D}\phi_{n}^{*}(\mathbf{r}_{1})\phi_{n}(\mathbf{r})/\sqrt{\mathcal{N}_{1}}$.
There exist $D-1$ normalized vectors orthogonal to $\vec{v}_{1}$
in a $D$-dimension Hilbert space, denoted as $\vec{u}_{n}$, $n=2,3,\cdots,D$
with $V\vec{u}_{n}=0\vec{u}_{n}$. These vectors, $\left\{ \vec{v}_{1},\vec{u}_{2},\cdots,\vec{u}_{D}\right\} $,
form a complete set of all the eigenvectors. This method for diagonalizing
rank-$1$ matrices has been used to obtain trilobite potentials between
a Rydberg atom and a ground state atom induced by high angular momentum
Rydberg electrons \citep{Greene2000PRL,Greene2001PRA}. With known
eigenstates and eigenenergies for $T$ and $H$, the many-body problem
for light fermions can be solved exactly. At zero temperature, the
ground state energy is the Fermi sea filling from low to high levels.
The inclusion of the interacting impurity shifts only one energy level
from the manifold by $g_{1}\mathcal{N}_{1}$, corresponding to the
spatial wave-function $\psi_{1}(\mathbf{r})$. The polaron energy,
defined as the total energy difference for $N$ fermions governed
by $T$ and $H$, is $E_{p}=g_{1}\mathcal{N}_{1}$.

For multiple impurities, the $V$ matrix can be written as $V=\sum_{s=1}^{S}V_{s}\equiv\sum_{s=1}^{S}g_{s}\mathcal{N}_{s}\vec{v}_{s}\vec{v}_{s}^{\dagger},$
where $\vec{v}_{s}^{\dagger}=\left[\phi_{1}(\mathbf{r}_{s}),\phi_{2}(\mathbf{r}_{s}),\cdots,\phi_{D}(\mathbf{r}_{s})\right]/\sqrt{\mathcal{N}_{s}}$
and $\mathcal{N}_{s}=\sum_{n=1}^{D}\left|\phi_{n}(\mathbf{r}_{s})\right|^{2}$.
This is a rank-$S$ matrix if $S<D$, with $S$ non-zero eigenvalues
$\lambda_{s}$ in ascending order. One can define orthogonal vectors
$\vec{u}_{n}$ for $n=S+1,S+2,\cdots,D$, orthogonal to all $\vec{v}_{s}$,
$s=1,2,\cdots,S$. These vectors satisfies $V\vec{u}_{n}=0\vec{u}_{n}$
and the corresponding eigenenergies remain unchanged in the manifold
despite the presence of impurities. The shifted eigenstates can be
solved in the Hilbert space spanned by the non-orthogonal $\vec{v}_{n}$
as a generalized eigenvalue problem $\tilde{V}\vec{c}_{s}=\lambda_{s}\tilde{S}\vec{c}_{s},$
where $\tilde{S}$ and $\tilde{V}$ are $S\times S$ matrices with
elements $\tilde{S}_{ij}=\vec{v}_{i}^{\dagger}\vec{v}_{j}$ and $\tilde{V}_{ij}=\vec{v}_{i}^{\dagger}V\vec{v}_{j}=\sum_{s=1}^{S}g_{s}\mathcal{N}_{s}\tilde{S}_{is}\tilde{S}_{sj}$,
respectively. The $S$ eigenvalues $\lambda_{s}$ correspond to the
$S$ eigenvectors $\vec{w}_{s}=\sum_{s^{\prime}=1}^{S}\vec{v}_{s^{\prime}}c_{s^{\prime}s}$,
where $c_{s^{\prime}s}$ is the $s^{\prime}$-th element of the column
vector $\vec{c}_{s}$. 

For the fully filled lowest manifold, the total energy change due
to $S$ impurities is $E_{S}=\sum_{s=1}^{S}\lambda_{s}={\rm Tr}[V]$,
since the remaining $D-S$ eigenvalues of $V$ are zero. Therefore,
$E_{S}=\sum_{s=1}^{S}{\rm Tr}[V_{s}]=\sum_{s=1}^{S}E_{p}^{(s)}$,
where $E_{p}^{(s)}=g_{s}\mathcal{N}_{s}$ is the single polaron energy
of the $s$-th impurity. We thus find that for the fully filled lowest
manifold, the interaction energy between any two impurities is zero,
so the polaron-polaron interactions are \emph{exactly }zero. This
conclusion holds for a non-fully-occupied lowest manifold if the number
of fermions in the degenerate manifold exceeds the number of impurities.

As a solid example, let us consider weakly interacting impurities
immersed in a 2D gas of non-interacting light fermions subjected to
a perpendicular uniform effective magnetic field in the $x$-$y$
plane with area $L_{x}L_{y}$. Using the Landau gauge $\mathbf{A}=Bx\hat{y}$,
the Landau Hamiltonian of the light fermions is
\begin{equation}
T=\frac{\left(\mathbf{p}-q\mathbf{A}\right)^{2}}{2m}=\frac{p^{2}}{2m}-\frac{qBp_{y}}{m}x+\frac{q^{2}B^{2}}{2m}x^{2},
\end{equation}
which results in highly degenerate Landau levels $E_{k_{y},n}=\hbar\omega_{c}\left(n+1/2\right)$,
recognizing the Hamiltonian as a harmonic oscillator centered at $x=k_{y}\ell_{B}^{2}$.
Here the cyclotron frequency $\omega_{c}=\left|qB\right|/m$ and the
magnetic length $\ell_{B}=\sqrt{\hbar/|qB|}$ are defined by the effective
magnetic field strength $B$, effective charge $q$, and mass $m$
of the light fermions. The momentum in the $y$-direction $p_{y}\equiv\hbar k_{y}$
is a good quantum number. 

Focusing on the case where only the lowest manifold is completely
filled, and assuming all impurity interactions are identical $g_{s}=g$,
the vector $\vec{v}_{s}$ for the impurity at $\mathbf{r}_{s}=\left(x_{s},y_{s}\right)$
is then a column vector indexed by $k_{y}$ as
\begin{equation}
\left(\vec{v}_{s}\right)_{k_{y}}=\frac{e^{ik_{y}y_{s}}}{\sqrt[4]{\pi\mathcal{N}_{s}\ell_{B}^{2}L_{y}^{2}}}\exp\left[-\frac{\left(x_{s}-k_{y}\ell_{B}^{2}\right)^{2}}{2\ell_{B}^{2}}\right].
\end{equation}
The normalization constant is $\mathcal{N}_{s}=\mathcal{N}\equiv1/(2\pi\ell_{B}^{2})$.
The polaron energy for a single impurity is therefore given by 

\begin{equation}
E_{p}=\frac{g}{2\pi\ell_{B}^{2}}.\label{eq:Ep_wk}
\end{equation}
To investigate the induced interaction between two impurities, we
place them at $(0,0)$ and $(0,R)$, respectively. The overlap can
be analytically obtained as $\vec{v}_{1}^{\dagger}v_{2}=\exp\left(-R^{2}/4\ell_{B}^{2}\right)$,
which gives the two non-zero eigenvalues of $V$:
\begin{equation}
\lambda_{\pm}(R)=\frac{g}{2\pi\ell_{B}^{2}}\left[1\pm\exp\left(-\frac{R^{2}}{4\ell_{B}^{2}}\right)\right],\label{eq:lambdapm_wk}
\end{equation}
corresponding to the symmetric and antisymmetric solutions $\vec{w}_{\pm}=\left(\vec{v}_{1}\pm\vec{v}_{2}\right)/\sqrt{2\pm2\exp\left(-R^{2}/4\ell_{B}^{2}\right)}$.
If the lowest manifold is completely filled, the polaron interaction
is $\lambda_{+}+\lambda_{-}-2E_{p}=0$, following the general consideration
stated earlier. It is convenient to define the contribution from each
state as $U_{\pm}\left(R\right)\equiv\lambda_{\pm}\left(R\right)-\lambda_{\pm}(\infty)$,
whose analytical expression reads as

\begin{equation}
U_{\pm}(R)=\pm\frac{g}{2\pi\ell_{B}^{2}}\exp\left(-\frac{R^{2}}{4\ell_{B}^{2}}\right).\label{eq:Urwk}
\end{equation}
It implies that the Born-Oppenheimer potential between two heavy impurities
is an attractive Gaussian potential, if there is only one light fermion
to mediate interactions.

\begin{figure}
\includegraphics[width=0.98\columnwidth]{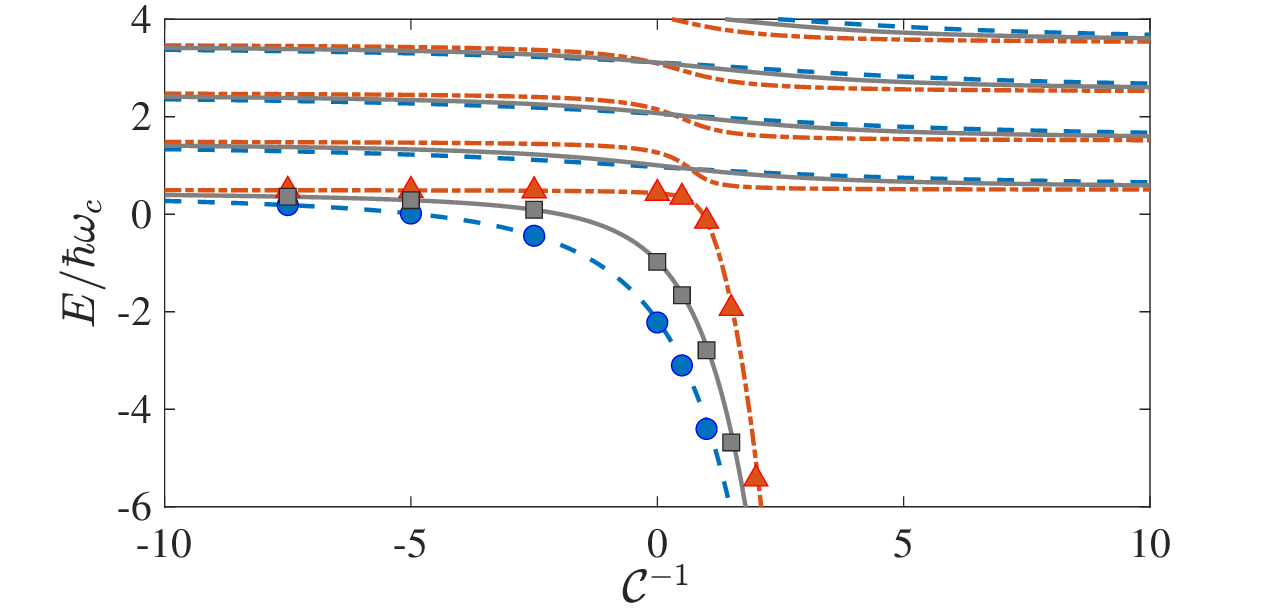}\caption{Single-particle energy levels shifted away from the degenerate manifolds
due to the presence of impurities, as a function of a regularized
coupling constant $\mathcal{C}^{-1}\equiv\log[\ell_{B}^{2}/(2a_{2D}^{2})]$.
Black solid curves represent single impurity, while blue dashed and
red dash-dotted curves represent $E_{+}$ and $E_{-}$ (see text below
Eq. (\ref{eq:TSeqSet}) for definition) of two impurities separated
by $R=0.5\ell_{B}$, respectively. Black squares, blue circles, and
red triangles indicate the results from numerical calculations with
2D van-der-Waals potentials (see text). \label{fig:EvsC}}
\end{figure}

\textbf{Strong impurity-fermion coupling }- For strong coupling between
impurities and fermions, we must go beyond the lowest manifold approximation.
We assume that there are $M$ manifolds coupled by the impurity-fermion
coupling, each with degeneracy $D_{i}>S$ and degenerate energy $\varepsilon_{i}$,
where $i=1,2,\cdots,M$. The dimension of the total Hamiltonian with
the $M$ manifolds is $G=\sum_{i=1}^{M}D_{i}$. For each manifold,
we define $\vec{v}_{s}^{(i)\dagger}=\left[\phi_{1}^{(i)}(\mathbf{r}_{s}),\phi_{2}^{(i)}(\mathbf{r}_{s}),\cdots,\phi_{D_{i}}^{(i)}(\mathbf{r}_{s})\right]/\sqrt{\mathcal{N}_{s}^{(i)}}$,
where the superscript $(i)$ denotes the $i$-th manifold. The corresponding
orthogonal vectors are denoted as $\vec{u}_{n}^{(i)}$, where $n=S,S+1,\cdots,D_{i}$.
Defining $S\times M$ vectors $\mathcal{V}_{s,i}^{\dagger}=[0_{D_{1}},0_{D_{2}},\cdots,\vec{v}_{s}^{(i)\dagger},\cdots,0_{D_{M}}]$
and similarly for $(G-S\times M)$ orthogonal vectors $\mathcal{U}_{n,i}$
in the full Hilbert space, the contact interaction matrix $V=\sum_{s}g_{s}\delta(\mathbf{r}-\mathbf{r}_{s})$
can be expanded as a block matrix with block elements as a $D_{i}\times D_{j}$
matrix given by $V_{ij}=\sum_{s}g_{s}\mathcal{V}_{s,i}\mathcal{V}_{s,j}^{\dagger}$.
We find $V\mathcal{U}_{n,i}=0$, indicating that $\mathcal{U}_{n,i}$
are eigenvectors of the Hamiltonian with the same eigenvalues as the
non-interacting Hamiltonian $T$. Thus, only $S$ states shift in
each manifold, and these shifted eigenstates can be solved as a generalized
eigenvalue problem in the Hilbert space spanned by the $S\times M$
unorthogonal vectors, which is much smaller than the full Hilbert
space of dimension $G$. While this problem generally requires large-scale
numerical solutions, interestingly we are able to derive a set of
much simplified equations to exactly solve the case of heavy and strongly
interacting impurities immersed in a Fermi sea of light fermions in
LLs (see Eq. (\ref{eq:TSeqSet}) below). In this case, the coupling
constant $g$ must be regularized by the 2D scattering length $a_{{\rm 2D}}$,
which naturally couples infinitely many manifolds.

\begin{figure}
\includegraphics[width=0.98\columnwidth]{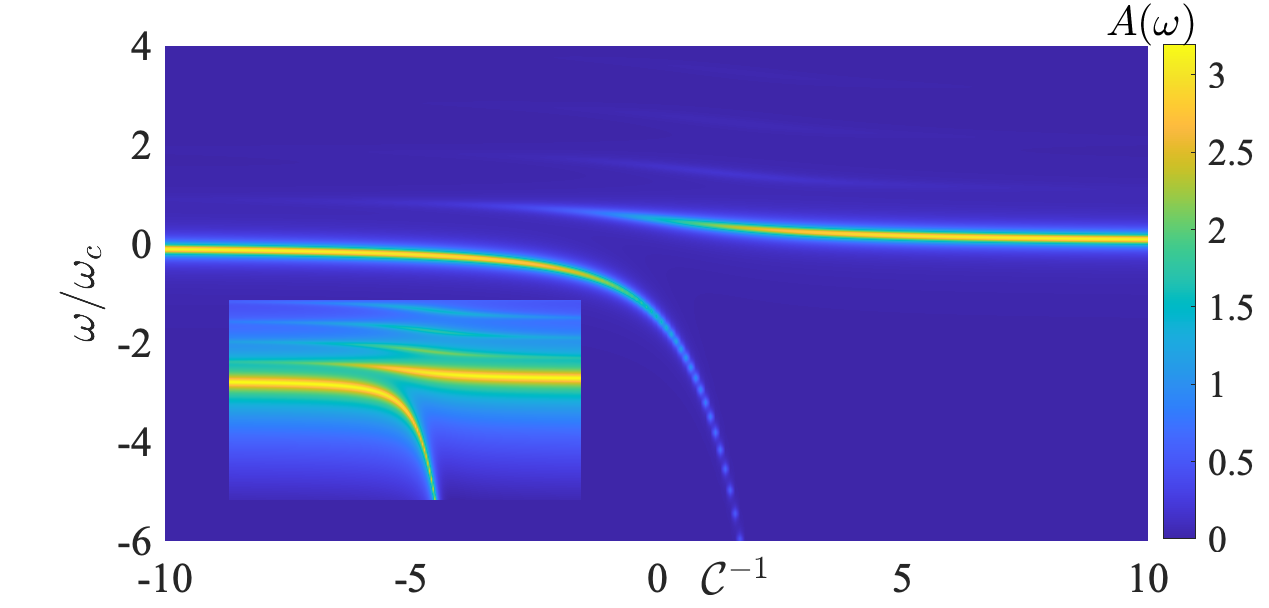}\caption{The spectral function $A(\omega)$ for a single infinitely heavy impurity
immersed in a Fermi sea with fully populated lowest LLs at zero temperature,
as a function of frequency $\omega$ and $\mathcal{C}^{-1}$. The
inset shows $A(\omega)$ in a logarithmic scale. All peaks have infinitesimally
small widths, with an artificial spectral broadening of $0.1\omega_{c}$
added for visibility. \label{fig:Aw}}
\end{figure}

To derive Eq. (\ref{eq:TSeqSet}), let us first examine the single
impurity case using the symmetric gauge $\mathbf{A}=\left(-By\hat{x}+Bx\hat{y}\right)/2$
and cylindrical coordinates centered at the impurity. Due to cylindrical
symmetry, the angular quantum number $m_{\ell}$ is a good quantum
number, and the non-interacting Hamiltonian is
\begin{equation}
\hat{T}=-\frac{\hbar^{2}}{2m}\frac{1}{r}\frac{\partial}{\partial r}\left(r\frac{\partial}{\partial r}\right)+\frac{m_{\ell}^{2}}{2mr^{2}}+\frac{m}{8}\omega_{c}^{2}r^{2}-\frac{\hbar\omega_{c}}{2}m_{\ell},
\end{equation}
equivalent to a 2D isotropic harmonic trap problem with an energy
shift of $-m_{\ell}\hbar\omega_{c}/2$. The eigenenergies are $E=\hbar\omega_{c}\left[n_{r}+1/2+\left(\left|m_{\ell}\right|-m_{\ell}\right)/2\right]$,
with $n_{r}=0,1,2,\cdots$ and $m_{\ell}=0,1,2,\cdots$. The degeneracy
pattern matches that of the Landau gauge, with each manifold having
one $s$-wave $(m_{\ell}=0)$. In the presence of an impurity at the
origin with contact interaction, only the $s$-wave state is affected,
agreeing with our analytical conclusion that only one state shifts
away from each manifold. 

Since $m_{\ell}=0$, the $s$-wave problem is equivalent to the case
in harmonic traps studied in Ref. \citep{Busch1998FP,Liu2010PRB}.
Solving the $s$-wave problem with the Bethe-Peierls boundary condition,
the radial Schr{\"o}dinger equation yields solutions that vanish at infinity:
$\Phi(\mathbf{r};\nu)=e^{-\xi/2}U(\nu,1,\xi)$, where $\nu=-E/\hbar\omega_{c}+1/2$,
$\xi=r^{2}/(2\ell_{B}^{2})$ and $U(\alpha,\beta,\xi)$ is the Tricomi
confluent hypergeometric function \citep{Liu2010PRB}. Near the origin,
\begin{equation}
e^{-\xi/2}U(\nu,1,\xi)\rightarrow\frac{\log(\sqrt{2}\ell_{B})-\psi[\nu]/2-\gamma-\log[r]}{\Gamma[\nu]/2},
\end{equation}
where $\gamma$ is the Euler--Mascheroni constant, $\psi[n,z]$ is
the $n$-th derivative of digamma function $\psi[z]$ at $z$ and
$\Gamma[z]$ denotes the gamma function. Applying the Bethe-Peierls
boundary condition by comparing with the free-space scattering wave
solution 
\begin{equation}
J_{0}(k\rho)-\tan(\delta)Y_{0}(k\rho)\rightarrow\frac{\log(2a_{2D})-\gamma-\log(r)}{\log(ka_{2D})}
\end{equation}
yields the transcendental equation:

\begin{equation}
\psi\left[\nu\right]=\mathcal{C}^{-1}\equiv\log\left(\frac{\ell_{B}^{2}}{2a_{2D}^{2}}\right),
\end{equation}
matching Ref. \citep{Busch1998FP}, with results shown as the black
solid curves of Fig. \ref{fig:EvsC}. If only the lowest LLs are fully
populated, the polaron energy can be obtained as $E_{p}=E-\hbar\omega_{c}/2=-\nu\hbar\omega_{c}$.
In the weak coupling regime, we have $\psi\left[\nu\right]\rightarrow\hbar\omega_{c}/E_{p}$
and $g\rightarrow2\pi\hbar^{2}\mathcal{C}/m$, leading to $E_{p}=g/2\pi\ell_{B}^{2}$,
which agrees with Eq. (\ref{eq:Ep_wk}). 

\begin{figure}
\includegraphics[width=0.98\columnwidth]{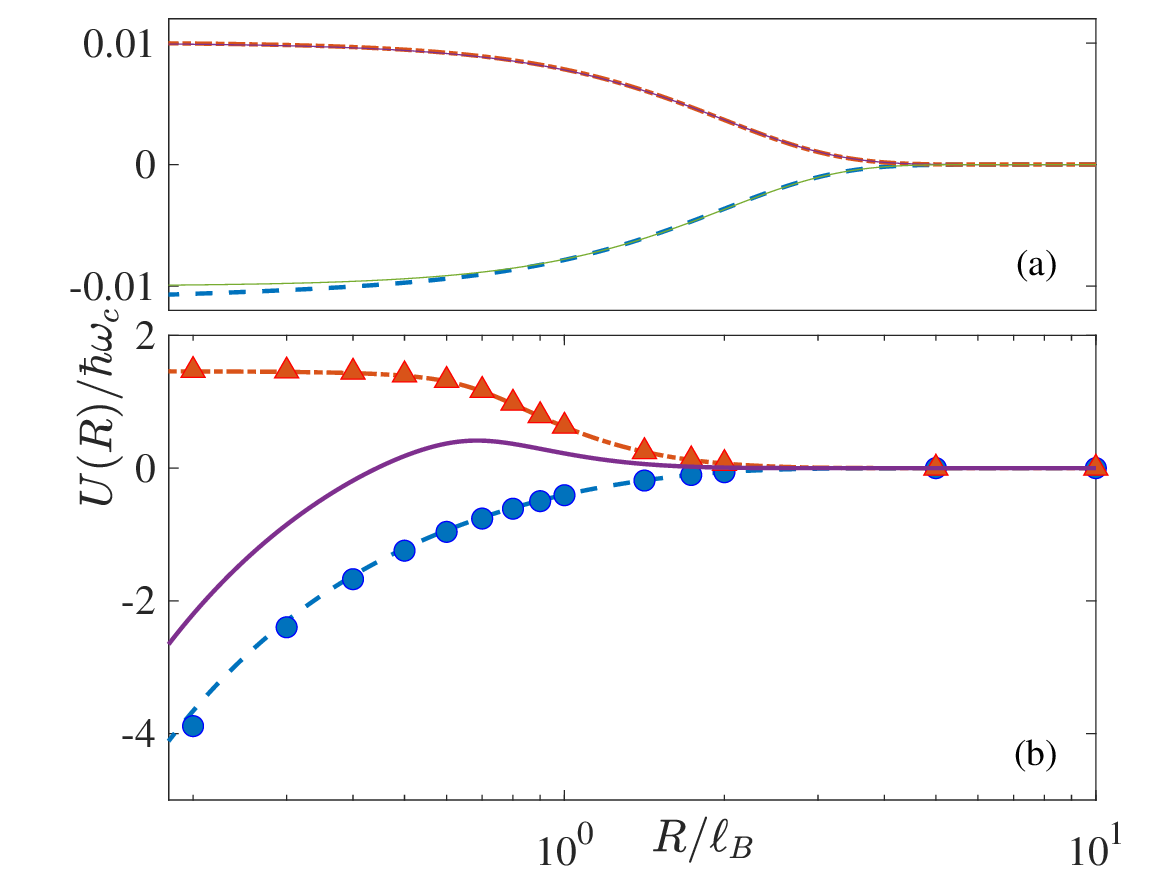}\caption{The potential between two heavy impurities, mediated by the light
fermions filled in the lowest LLs, with (a) weak impurity-fermion
coupling $\mathcal{C}=0.01$ and (b) strong coupling $\mathcal{C}^{-1}\rightarrow0$.
The symmetric and antisymmetric state potentials are shown by the
blue dashed and red dash-dotted curves, respectively. In (a), the
potentials are compared with the Gaussian potentials (thin solid curves)
from Eq. (\ref{eq:Urwk}) in the lowest manifold approximation. In
(b), the polaron interaction $U_{+}(R)+U_{-}(R)$ for fully populated
lowest LLs is depicted by the purple solid curve. \label{fig:Ur}}
\end{figure}

Not only the polaron energy, but also the entire polaron spectrum
can be exactly and analytically determined via the functional determinant
approach \citep{Leonid1996JMathPhys,Klich2003Book,Schonhammer2007PRB,JiaPRL2022,Jia2022PRA,Wang2022Review}.
At zero temperature, with only the lowest LLs fully populated, we
have $A(\omega)={\rm Re}\int_{0}^{\infty}S(t)e^{i\omega t}dt/\pi$,
where ${\rm Re}$ denotes the real part, and $S(t)=\sum_{\nu}e^{-i\nu\omega_{c}t}/\left(\nu^{2}\psi[1,\nu]\right)$,
with proofs given in Appendix \ref{sec:App2}. The corresponding spectral
function, $A(\omega)$, shown in Fig. \ref{fig:Aw}, can be measured
via polaron spectroscopy \citep{Cetina2016Science}. Notably, the
polaron exhibits multiple branches in positive frequency (better illustrated
in the inset with logarithmic scale), all with infinitesimal widths.
This differs significantly from conventional repulsive polarons and
possibly results from the suppression of particle-hole excitations
due to the energy gaps between different LL manifolds.

For multiple impurities, inspired by the singularity regularization
using a single set of harmonic states around the impurity \citep{Busch1998FP},
we express the total wave function as a superposition of single impurity
states: $\Psi(\mathbf{r};E)\sim\sum_{s}c_{s}\Phi_{s}(\mathbf{r}-\mathbf{r}_{s};\nu)$,
where $\Phi_{s}(\mathbf{r}-\mathbf{r}_{s};\nu)=e^{i\varphi_{s}}e^{-\xi_{s}/2}U\left(\nu,1,\xi_{s}\right)$,
$\xi_{s}=\left(\mathbf{r}-\mathbf{r}_{s}\right)^{2}/(2\ell_{B}^{2})$,
and $\varphi_{s}=|\mathbf{r}_{s}\times\mathbf{r}|/(2\ell_{B}^{2})$
is a phase compensating the gauge difference between the origin and
$\mathbf{r}_{s}$ (see Appendix \ref{sec:App3} for details). $\Psi(\mathbf{r};E)$
is an eigenstate with eigenenergy $\nu=-E/\hbar\omega_{c}+1/2$ except
at $\mathbf{r}_{s}$. Imposing the Bethe-Peierls boundary conditions
at $\mathbf{r}\rightarrow\mathbf{r}_{s}$ gives $S$ transcendental
equations,
\begin{equation}
\left(\psi\left[\nu\right]-\mathcal{C}_{s}^{-1}\right)c_{s}=\Gamma[\nu]\sum_{s^{\prime}\ne s}\Phi(\mathbf{r}_{s}-\mathbf{r}_{s^{\prime}};\nu)c_{s^{\prime}}.\label{eq:TSeqSet}
\end{equation}
The solutions of these equations provide the eigenenergies. Figure
\ref{fig:EvsC} shows the results for two impurities with the same
$a_{{\rm 2D}}$ separated by a distance $R=0.5\ell_{B}$. We find
that there are exactly two solutions shifted away in each manifold,
agreeing with our previous analysis. The lowest two eigenenergies
$E_{\pm}$ correspond to the symmetric and antisymmetric wavefunctions
$\Psi_{\pm}(\mathbf{r})\sim\Phi(\mathbf{r}+\mathbf{R}/2)\pm\Phi(\mathbf{r}-\mathbf{R}/2)$. 

Next, we investigate the distance dependency of the mediated interaction
between two impurities. Figure \ref{fig:Ur} (a) shows the potential
$U_{\pm}(R)\equiv E_{\pm}(R)-E_{\pm}(\infty)$ in the weak-coupling
regime ($\mathcal{C}=0.01$), which matches well with the Gaussian
potentials from Eq. (\ref{eq:Urwk}) in the lowest manifold approximation.
Fig. \ref{fig:Ur} (b) presents the potentials in the strong-coupling
regime ($\mathcal{C}^{-1}\rightarrow0$). The asymptotic behavior
of the symmetric state potential $U_{+}$ is obtained via fitting:
$U_{+}(R)\rightarrow-\exp(-\tilde{c}R^{2}/\ell_{B}^{2})$ with coefficient
$\tilde{c}\approx0.275$ for $R\gg\ell_{B}$, and $U_{+}(R)\rightarrow-\tilde{d}\ell_{B}/R$
with $\tilde{d}\approx0.8$ for $R\ll\ell_{B}$. For a single light
fermion, $U_{+}(R)$ describes the Born-Oppenheimer potential between
two heavy impurities. The antisymmetric state potential $U_{-}(R)$
shows a repulsive tail $\exp(-\tilde{c}R^{2}/\ell_{B}^{2})$ at large
distance and goes to a constant at short range. For a fully populated
manifold, the induced polaron interaction $U_{+}(R)+U_{-}(R)$ is
essentially zero for $R\gg\ell_{B}$ and exhibits a strongly attractive
$1/R$ behavior at short-range, with a potential barrier at $R\approx0.685\ell_{B}$.

To verify our analytical results, we finally conduct direct, large-scale
numerical calculations using 2D van-der-Waals potentials with a length
scale $l_{{\rm vdW}}\ll\ell_{B}$ (see Appendix \ref{sec:App4} for
details). The numerical results, shown in Fig. \ref{fig:EvsC} and
Fig. \ref{fig:Ur} (b) as symbols, align perfectly with our analytical
results, except for slight deviations at very small $R\sim l_{{\rm vdW}}$,
as expected.

\textbf{Conclusions} - We have derived analytical expressions for
the induced interactions and polaron energies of heavy impurities
mediated by light fermions with high energy-degeneracy at weak impurity-fermion
coupling. We have also exactly solved the mediated polaron-polaron
interactions by numerical calculations for strong impurity-fermion
coupling. The short-range attractive polaron-polaron potential predicted
at the strong coupling implies that impurities within $\ell_{B}$
will bind into molecules, while impurities at distances greater than
$\ell_{B}$ will not interact at all. Our exact results can be directly
examined with fermionic systems in the quantum Hall regime, to be
realized in cold-atom experiments \citep{Zhou2023Science} or already
observed in 2D condensed matter setups \citep{Imamoglu2018PRL}. Furthermore,
our general analysis of polaron-polaron interactions with highly degenerate
levels is widely applicable to different situations including bosons
as exchanged particles. Although bosonic statistics ensures that ideal
or weakly interacting bosons mostly fill the lowest energy level at
zero temperature, our results might be beneficial for engineering
LL state occupations in recent experiments involving rotating Bose-Einstein
condensates \citep{CornellPRL2004,Spielman2009Nature,Zwierlein2021Science,Zwierlein2022Nature}
with point-like localized potentials induced by lasers.
\begin{acknowledgments}
This research was supported by the Australian Research Council's (ARC)
Discovery Program, Grants Nos. FT230100229 (J.W.), DP240100248 (X.-J.L.),
and DP240101590 (H.H.).
\end{acknowledgments}

\appendix

\section{Repulsively weakly interacting impurities\label{sec:App1}}

In the main text, we have concluded that for attractive and weak impurity-fermion
coupling, the heavy polaron energy is $E_{p}=g_{1}\mathcal{N}_{1}$,
if only the lowest manifold is involved. When $g_{1}>0$, the shifted
energy level is higher than the degenerate manifold, and hence the
polaron energy is zero, if the manifold is not fully populated. In
this case, it turns out to be more convenient to consider holes instead
of particles (i.e., the light fermions), that is, if there is a particle
state is not occupied in the manifold, we understand it as an occupied
hole state. In other words, the polaron energy is zero, if there is
at least one hole in the manifold. 

For multiple $S$ impurities, the polaron energy is again exactly
zero if the number of holes in the manifold is larger than the number
of impurities. This observation is parallel to the the conclusion
that we have made to the attractive coupling case in the main text,
i.e., the polaron energy is precisely zero, when the light fermionic
particles outnumber the impurities. Moreover, for the case of two
heavy impurities separated by a relative distance $R$ in the lowest
LLs with identical interaction $g>0$, the polaron-polaron interaction
mediated by a single hole is an attractive Gaussian potential, 
\begin{equation}
U_{-}(R)=-\frac{g}{2\pi\ell_{B}^{2}}\exp\left(-\frac{R^{2}}{4\ell_{B}^{2}}\right).
\end{equation}

\section{Functional Determinant Approach\label{sec:App2}}

For an infinitely heavy impurity immersed in a non-interacting Fermi
sea, the polaron spectroscopy can be obtained via an exact method,
namely the functional determinant approach \citep{Leonid1996JMathPhys,Klich2003Book,Schonhammer2007PRB,JiaPRL2022,Jia2022PRA,Wang2022Review}.
In such an approach, the spectral function is given by $A(\omega)={\rm Re}\int_{0}^{\infty}S(t)e^{i\omega t}dt/\pi$,
where the overlapping function at zero temperature is given by
\begin{equation}
S(t)=\det\left[\mathbf{1}-\hat{n}+R(t)\hat{n}\right],
\end{equation}
with $\mathbf{1}$ being an identity matrix, and $\hat{n}$ being
a diagonal matrix with matrix elements being the occupation number.
The matrix 
\begin{equation}
R(t)=e^{iTt/\hbar}e^{-iHt/\hbar}
\end{equation}
are determined by the single-particle Hamiltonian $T$ and $H$, and
implies that the unshifted eigenstates do not contribute to $S(t)$.
For the case where only the lowest LLs are fully populated, 
\begin{equation}
S(t)=\sum_{\nu}\left|\left\langle n=0\right.\left|\nu\right\rangle \right|^{2}e^{i\nu\omega_{c}t}.
\end{equation}
Here $\left|n=0\right\rangle $ denotes the lowest $s$-wave 2D harmonic
oscillator, and $\left|\nu\right\rangle $ denote the shifted $s$-wave
states due to the presence of the impurity with corresponding eigenenergies
$\nu=-E/\hbar\omega_{c}+1/2$. From Ref. \citep{Busch1998FP}, we
have
\begin{equation}
\left\langle n\right.\left|\nu\right\rangle =\frac{A_{\nu}\varphi_{n}(0)}{E-E_{n}},
\end{equation}
where $\varphi_{n}(0)=1/(2\sqrt{2}\ell_{B})$ is the $s$-wave 2D
harmonic oscillator wavefunction at the origin, and $E_{n}=(n+1/2)\hbar\omega_{c}$.
The coefficient $A_{\nu}$ can be obtained from the normalization
condition $\sum_{n=1}^{\infty}\left|\left\langle n\right.\left|\nu\right\rangle \right|^{2}=1$.
We find that
\begin{equation}
\left\langle n\right.\left|\nu\right\rangle =-\frac{1}{\left(\nu+n\right)\sqrt{\psi[1,\nu]}},
\end{equation}
and consequently,
\begin{equation}
S(t)=\sum_{\nu}\frac{e^{-i\nu\omega_{c}t}}{\nu^{2}\psi[1,\nu]}
\end{equation}
and
\begin{equation}
A(\omega)={\rm Re}\sum_{\nu}\frac{1/\left(\nu^{2}\psi[1,\nu]\right)}{\omega+i\eta-\nu\omega_{c}},
\end{equation}
where $\psi[n,z]$ is the $n$-th derivative of digamma function $\psi[z]$,
and $\eta=0^{+}$ is an infinitesimal positive number.

\section{Local phase of wavefunction\label{sec:App3}}

In the main text, we have solved the problem for a single impurity
at the origin with an interacting potential $V(x,y)$. The eigenstate
$\Phi(x,y)$ satisfies the Schr{\"o}dinger equation $\hat{H}\Phi(x,y)=E\Phi(x,y)$,
where the Hamiltonian in the symmetric gauge is:
\begin{widetext}
\begin{equation}
\hat{H}=\frac{\left(\hat{p}_{x}+qBy/2\right)^{2}}{2m}+\frac{\left(\hat{p}_{y}-qBx/2\right)^{2}}{2m}+V(x,y).
\end{equation}
If the impurity is shifted to $(x_{1},y_{1})$, the potential becomes
$V(x-x_{1},y-y_{1})$. The eigenenergy remains the same, and the probability
density of the eigenstate becomes $\left|\Phi(x-x_{1},y-y_{1})\right|^{2}$,
requiring a local phase to be determined in the wavefunction.

By transforming the coordinates $x\rightarrow x-x_{1}$ and $y\rightarrow y-y_{1}$,
the Hamiltonian becomes 
\begin{equation}
\hat{H}\rightarrow\frac{\left(\hat{p}_{x}+qBy/2-qBy_{1}/2\right)^{2}}{2m}+\frac{\left(\hat{p}_{y}-qBx/2+qBx_{1}/2\right)^{2}}{2m}+V(x-x_{1},y-y_{1}).
\end{equation}
Introducing $\hat{p}_{x}\rightarrow\hat{p}_{x}+qBy_{1}/2$ and $\hat{p}_{y}\rightarrow\hat{p}_{y}-qBx_{1}/2$,
the Hamiltonian converts to
\begin{equation}
\hat{H}\rightarrow\frac{\left(\hat{p}_{x}+qBy/2\right)^{2}}{2m}+\frac{\left(\hat{p}_{y}-qBx/2\right)^{2}}{2m}+V(x-x_{1},y-y_{1}),
\end{equation}
which is the desired Hamiltonian for an impurity at $(x_{1},y_{1})$.
This transformation can be achieved by a gauge transformation: $\Phi(x-x_{1},y-y_{1})\rightarrow e^{iqB\left(xy_{1}-x_{1}y\right)/2\hbar}\Phi(x-x_{1},y-y_{1})$.
As a result, the wavefunction acquires a local phase $e^{i\varphi_{1}}$,
where
\begin{equation}
\varphi_{1}=\frac{\left|\mathbf{r}_{1}\times\mathbf{r}\right|}{2\ell_{B}^{2}},
\end{equation}
and $\ell_{B}^{2}=\hbar/\left|qB\right|$. 
\end{widetext}

\section{Large-scale numerical Investigation\label{sec:App4}}

To verify our analytical expressions, we have performed direct numerical
calculations using a 2D Lennard-Jones potential
\begin{equation}
V(x,y)=-\frac{C_{6}}{r^{6}}\left(1-\frac{\lambda_{0}^{6}}{r^{6}}\right),
\end{equation}
where the van der Waals length $l_{{\rm vdW}}=\sqrt[4]{2mC_{6}/\hbar^{2}}/2$
characterizes the potential range, and the short-range parameter $\lambda_{0}$
is adjusted to achieve the desired 2D scattering length $a_{{\rm 2D}}$. 

For numerical calculations of Landau levels in the presence of impurities,
we adopted the Landau gauge and diagonalized the Hamiltonian in a
large square box $L_{{\rm Box}}\times L_{{\rm Box}}$ in Cartesian
coordinates, i.e., $L_{x}=L_{y}=L_{\textrm{Box}}$. Under the condition
$l_{{\rm vdW}}\ll\ell_{B}\ll L_{{\rm Box}}$, the numerical results
should match our analytical findings for impurities with contact interactions
in free space. In practise, we set $L_{{\rm Box}}=100\times\sqrt{2}\ell_{B}=10^{4}l_{{\rm vdW}}$.
Convergence was tested for box size, grid discretization, and the
number of deep bound states supported by the Lennard-Jones potential. 

\bibliography{RefLandauPolaron}

\end{document}